\documentclass[twocolumn,showpacs,preprintnumbers,superscriptaddress,amsmath,amssymb,pre]{revtex4-1}

\usepackage{booktabs,threeparttable}
\usepackage{graphicx}
\usepackage{dcolumn}
\usepackage{bm}
\usepackage{color}
\usepackage{multirow}
\usepackage{longtable}

\begin{document}

\title{Calling Patterns in Human Communication Dynamics}

\author{Zhi-Qiang Jiang}
\email{zqjiang@ecust.edu.cn}
\affiliation{School of Business, School of Science, and Research Center for Econophysics, East China University of Science and Technology, Shanghai 200237, China}
 
\author{Wen-Jie Xie}
\affiliation{School of Business, School of Science, and Research Center for Econophysics, East China University of Science and Technology, Shanghai 200237, China}
 
\author{Ming-Xia Li}
\affiliation{School of Business, School of Science, and Research Center for Econophysics, East China University of Science and Technology, Shanghai 200237, China}
  
\author{Boris Podobnik}
\affiliation{Center for Polymer Studies and Department of Physics, Boston University, Boston, Massachusetts 02215, USA}
\affiliation{Zagreb School of Economics and Management, 10000 Zagreb, Croatia}

\author{Wei-Xing Zhou}
\email{wxzhou@ecust.edu.cn}
\affiliation{School of Business, School of Science, and Research Center for Econophysics, East China University of Science and Technology, Shanghai 200237, China}

\author{H. Eugene Stanley}
\email{hes@bu.edu}
\affiliation{Center for Polymer Studies and Department of Physics, Boston University, Boston, Massachusetts 02215, USA}



\begin{abstract}
Modern technologies not only provide a variety of communication modes, e.g., texting, cellphone conversation, and online instant messaging, but they also provide detailed electronic traces of these communications between individuals. These electronic traces indicate that the interactions occur in temporal bursts. Here, we study the inter-call durations of the 100,000 most-active cellphone users of a Chinese mobile phone operator. We confirm that the inter-call durations follow a power-law distribution with an exponential cutoff at the population level but find differences when focusing on individual users. We apply statistical tests at the individual level and find that the inter-call durations follow a power-law distribution for only 3460 individuals (3.46\%). The inter-call durations for the majority (73.34\%) follow a Weibull distribution.  We quantify individual users using three measures: out-degree, percentage of outgoing calls, and communication diversity. We find that the cellphone users with a power-law duration distribution fall into three anomalous clusters: robot-based callers, telecom frauds, and telephone sales. This information is of interest to both academics and practitioners, mobile telecom operator in particular. In contrast, the individual users with a Weibull duration distribution form the fourth cluster of ordinary cellphone users.  We also discover more information about the calling patterns of these four clusters, e.g., the probability that a user will call the $c_r$-th most contact and the probability distribution of burst sizes. Our findings may enable a more detailed analysis of the huge body of data contained in the logs of massive users.

\end{abstract}


\maketitle

Understanding the temporal patterns of individual human
interactions is essential in managing information spreading and in tracking
social contagion.  Human interactions, e.g., cellphone conversations and
emails, leave electronic traces that allow the tracking of human
interactions from the perspective of either static complex networks
\cite{Eckmann-Moses-Sergi-2004-PNAS,Palla-Barabasi-Vicsek-2007-Nature,Onnela-Saramaki-Hyvonen-Szabo-Menezes-Kaski-Barabasi-Kertesz-2007-NJP,Onnela-Saramaki-Hyvonen-Szabo-Lazer-Kaski-Kertesz-Barabasi-2007-PNAS,Jo-Pan-Kaski-2011-PLoS1,Kumpula-Onnela-Saramaki-Kaski-Kertesz-2007-PRL}
or human dynamics \cite{Barabasi-2005-Nature}. Because static networks
only describe sequences of instantaneous interacting links, temporal
networks in which the temporal patterns of interacting activities for
each node are recorded have recently received a considerable amount of
research interest
\cite{Holme-Saramaki-2012-PR,Pan-Saramaki-2011-PRE}. Investigations of
inter-event intervals between two consecutive interacting actions, such
as email communications
\cite{Barabasi-2005-Nature,Malmgren-Stouffer-Motter-Amaral-2008-PNAS},
short-message correspondences
\cite{Hong-Han-Zhou-Wang-2009-CPL,Wu-Zhou-Xiao-Kurths-Schellnhuber-2010-PNAS,Zhao-Xia-Shang-Zhou-2011-CPL},
cellphone conservations
\cite{Candia-Gonzalez-Wang-Schoenharl-Madey-Barabasi-2008-JPAMT,Karsai-Kaski-Barabasi-Kertesz-2012-SR},
and letter correspondences
\cite{Oliveira-Barabasi-2005-Nature,Li-Zhang-Zhou-2008-PA,Malmgren-Stouffer-Campanharo-Amaral-2009-Science},
indicate that human interactions have non-Poissonian characteristics.
Previous studies were conducted either on aggregate samples
\cite{Candia-Gonzalez-Wang-Schoenharl-Madey-Barabasi-2008-JPAMT,Karsai-Kaski-Barabasi-Kertesz-2012-SR,Karsai-Kaski-Kertesz-2012-PLoS1}
or on a small group of selected individuals
\cite{Barabasi-2005-Nature,Oliveira-Barabasi-2005-Nature,Li-Zhang-Zhou-2008-PA,Malmgren-Stouffer-Motter-Amaral-2008-PNAS,Malmgren-Stouffer-Campanharo-Amaral-2009-Science,Hong-Han-Zhou-Wang-2009-CPL,Wu-Zhou-Xiao-Kurths-Schellnhuber-2010-PNAS},
but the communication behavior of individuals is not well understood.

We study the complete voice information for cellphone users supplied by
a Chinese cellphone operator and study the inter-event time between two
consecutive outgoing calls (inter-call duration). Our studies are performed
at both the individual and group levels.  To ensure
better statistics, the top 100,000 cellphone users with the largest
number of outgoing calls are chosen as our data sample, each having more
than 997 outgoing calls. We propose a bottom-up approach to investigate
individual cellphone communication dynamics: (1) finding the functional
form of the distribution of each individual's intercall durations, (2) grouping
individuals with the same distribution, and (3) understanding the
calling patterns for each group.  We apply an automatic fitting
technology to each mobile phone user, and filter out two groups of users
according to their inter-call duration distributions. One group is
comprised of individuals with a power-law duration distribution (3,464
individuals) and the other is comprised of individuals with a Weibull
duration distribution (73,339 individuals). We demonstrate that the two
groups exhibit different calling patterns and that the individuals from
the power-law group exhibit anomalous communication behaviors (e.g., the group
includes individuals sending spam).

\bigskip
{\textbf{Results}}

{\textbf{Distribution at the population level.}}
There are 5,921,696 different individuals in our data set (see the data
description in Materials and Methods). For each individual, we estimate
the intraday inter-call durations ($d$ seconds) (see definition of
intraday inter-call durations in Materials and Methods) and we find
that 4,635,536 individuals have non-empty intraday durations ($n_d >
0$), which we consider one unique sample when we investigate the
distribution at the population level.  To this end we also analyze the
aggregate level where the data comprise only the durations of the top
100,000 individuals.

\begin{figure}[htbp]
 \centering
 \includegraphics[width=8cm]{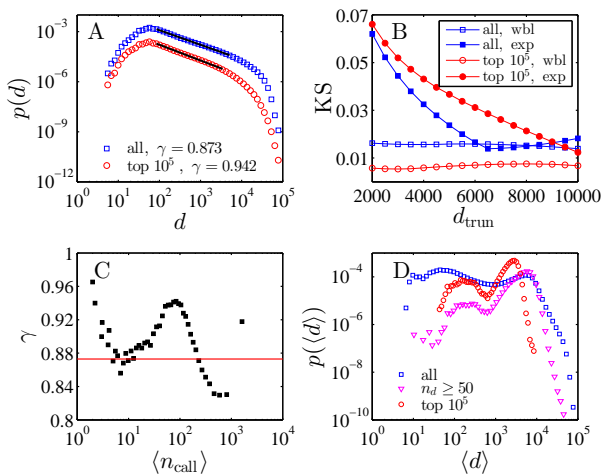}
 \caption{\label{Fig1:Agg:Durations} Probability distribution of the
   inter-call durations ($d$ seconds). ({\textit{A}}) Distribution of inter-call
   durations at the population level. The circle markers are shifted
   vertically by a factor of 0.1 for better visibility. ({\textit{B}}) Plots of the
   statistic $\rm{KS}$ with respect to the truncated value
   $d_{\rm{trun}}$ for Weibull and exponential distribution. ({\textit{C}}) Plots
   of the power-law exponent $\gamma$ with respect to the average number
   of outgoing calls $\langle n_{\rm{call}} \rangle$ for different
   groups. The red line stands for the power-law exponent of the whole
   sample. ({\textit{D}}) Probability distribution of mean inter-call durations for
   different samples of individuals.}
\end{figure}

Figure~\ref{Fig1:Agg:Durations}{\textit{A}} shows the empirical distributions of the
two samples of aggregate data. Both curves exhibit excellent power
law behaviors in the range of [80,2000] seconds. We apply the
least-square method, and find that a linear fit gives the power-law
exponent $\gamma_{\rm{all}} = 0.873$ for all the individuals and
$\gamma_{{\rm{top}}~10^5} = 0.942$ for the top 100,000 individuals,
respectively. We compare the empirical distributions obtained from our
dataset with the empirical distributions of inter-call durations based
on a different dataset provided by a European cellphone operator
\cite{Candia-Gonzalez-Wang-Schoenharl-Madey-Barabasi-2008-JPAMT} (see
also the supplementary information of
Ref.~\cite{Gonzalez-Hidalgo-Barabasi-2008-Nature}), and note that the
empirical distributions of both datasets share very similar patterns for
$d \le 10^5$, where only intraday inter-call durations are taken into
consideration. The reported power-law exponent $\gamma = 0.9$ in
Ref.~\cite{Gonzalez-Hidalgo-Barabasi-2008-Nature} is approximately equal
to the estimated exponents $\gamma_{\rm{all}}$ shown in
Fig.~\ref{Fig1:Agg:Durations}{\textit{A}}.  A similar functional form with a
power-law exponent $\gamma = 0.7$ is also reported in
Ref.~\cite{Karsai-Kaski-Barabasi-Kertesz-2012-SR} for inter-call
durations smaller than $10^5$. This similarity is further consolidated by
fitting the empirical duration distributions by means of a formula of power-law with an exponential
cutoff.

Figure \ref{Fig1:Agg:Durations}{\textit{A}} shows a clear deviation from the
power-law distribution in the tails of both curves, which is usually
interpreted as an exponential cut-off. To test which distribution better
fits the data, we apply Kolmogorov-Smirnov (KS) statistics by means of
which the smaller the value, the better the fit.  We set the truncation
value at 2000 seconds and find that for all individuals the tail is
better fit by the Weibull distribution ($\rm{KS} = 0.016$) than by the
exponential distribution ($\rm{KS} = 0.063$). Similarly, for the top
$10^5$ individuals the Weibull distribution ($\rm{KS} = 0.006$) also
fits the data better than the exponential distribution ($\rm{KS} =
0.066$).  Figure~\ref{Fig1:Agg:Durations}{\textit{B}} for varying truncation values
shows a plot of $\rm{KS}$ as a function of $d_{\rm{trun}}$. The
$\rm{KS}$ statistic displays a more stable behavior for the Weibull fit
than for the exponential fit, indicating that the Weibull distribution
is better able to capture the tail behavior than the exponential
distribution.

We further divide the sequence of individuals according to the number of
outgoing calls into 46 groups, sorted in ascending order.  The first
group comprises 135,536 individuals and the remaining 45 groups each
comprise 100,000 individuals. We calculate the empirical distributions
of the aggregate inter-call durations for each group and find that all
the distributions share patterns similar to those shown in
Fig.~\ref{Fig1:Agg:Durations}{\textit{A}}. Figure~\ref{Fig1:Agg:Durations}{\textit{C}} shows a
plot of the estimated power-law exponents with respect to the average
number of outgoing calls.  All the power-law exponents are lower than 1
and the mean value is $0.896 \pm 0.033$.

Figure~\ref{Fig1:Agg:Durations}{\textit{D}} shows the probability distributions of
the individual average inter-call durations calculated for (i) all the
individuals, (ii) the individuals with $n_d \ge 50$, and (iii) the top
100,000 individuals, respectively.  All three curves exhibit an
approximate $M$-shape characterized by two peaks.  For the sample of all individuals, there is a
large number of low-frequency individuals who do not use a
cellphone regularly.  The influence of these low-frequency callers is
eliminated in the distributional curve of $n_d \ge 50$. We compare this
distributional curve with the distribution of the top 100,000
individuals and find that they exhibit the same $M$-shape with a central
valley at approximately $d = 650$, strongly indicating the presence of
two groups of individuals possessing different calling patterns across
the sample.  One group is of individuals that have low average
inter-call duration values, indicating a high frequency of outgoing
calls, and the other is of individuals that have large average
inter-call duration values, indicating a relatively low frequency of
outgoing calls.  We will later demonstrate that the group with a high
frequency of outgoing calls is dominated by individuals with a power-law
duration distribution and that the group with a low frequency of
outgoing calls is dominated by the individuals with a Weibull duration
distribution.

\begin{figure}[htbp]
 \centering
 \includegraphics[width=8cm]{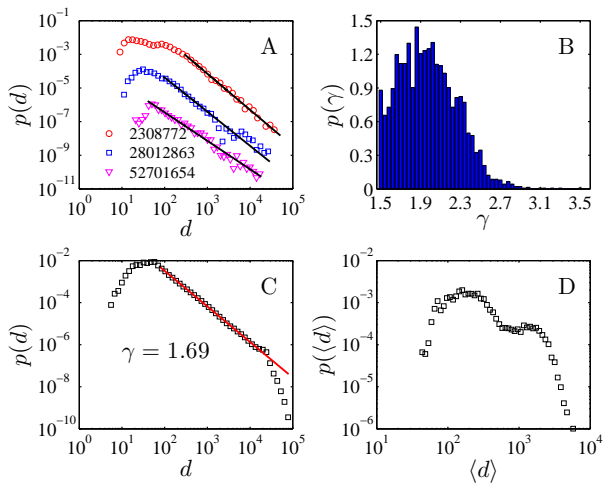}
  \caption{\label{Fig2:PowerLawGroup} Classified results of individuals with power-law distributions of intraday inter-call durations. ({\textit{A}})   Probability distributions of intraday durations for three randomly chosen individuals. The markers of individual 28012863 and 52701654 are shifted vertically by a factor of $10^{-2}$ and $10^{-4}$ for better visibility. The solid lines are the best maximum likelihood estimation (MLE) fit to the power-law distributions, which gives the power-law exponents $\gamma = 2.15$, $\gamma = 2.03$, and $\gamma = 1.69$ for individual 2308772, 28012863, and 52701654, respectively. ({\textit{B}}) Distribution of the estimated power-law exponents. ({\textit{C}}) Probability distribution of collective inter-call durations by aggregating the durations of different individuals from the power-law group as one sample. The solid line is the best fit to the data by means of the least-square method, which gives an estimation of power-law exponent $\gamma =1.69$. ({\textit{D}}) Probability distribution of the mean inter-call durations for the power-law group.}
\end{figure}

{\textbf{Classification of cellphone users.}}
According to the above analysis at the aggregate level, we propose to
classify the individuals according to their duration
distributions. Motivated by
Ref.~\cite{Clauset-Shalizi-Newman-2009-SIAMR}, but here for each
individual cellphone user, because we are focusing on the tail of the
distribution we assume the candidate duration distributions to be
left-truncated and we assign each of them a distribution that is either
power law or Weibull. We estimate the truncation value $d_{\min}$
associated with distribution parameters by finding the minimum KS
statistic. We then apply statistical tests to check the significance of
the fitting parameters (see fitting distributions and statistical tests
in Materials and Methods).  Finally, based on statistical tests, we find
that there are 3,464 individuals whose intraday durations follow a
power-law distribution and 73,339 individuals whose intraday durations
follow a Weibull distribution (see determining the distribution form in
Materials and Methods).

Figure~\ref{Fig2:PowerLawGroup}{\textit{A}} shows that the empirical duration
distributions for three randomly chosen individuals (2308772, 28012863,
and 52701654) whose intraday
inter-call durations follow a power-law distribution. The solid lines
correspond to the power-law fits with power-law exponents $\gamma =
2.15$, $\gamma = 2.03$, and $\gamma = 1.69$ for individuals 2308772,
28012863, and 52701654, respectively. Figure~\ref{Fig2:PowerLawGroup}{\textit{B}}
plots the distribution of the estimated power-law exponents $\gamma$ for
all individuals with an intraday inter-call duration that follows a
power-law distribution and finds that none of the power-law exponents
are lower than 1.5. This is in sharp contrast to the power-law exponents
lower than 1 that we found for the aggregate durations in
Fig.~\ref{Fig1:Agg:Durations}{\textit{C}}. Note that there is a large fraction of
individuals whose power-law exponents are between 1 and 3, which are the
characteristic values for the L{\'e}vy regime $(1,3)$. Note that the
exponent 2 corresponds to the famous Zipf law. Having all the power-law
exponents, we calculate the mean $\langle \gamma \rangle = 2.00 \pm
0.32$.

\begin{figure}[htbp]
 \centering
 \includegraphics[width=8cm]{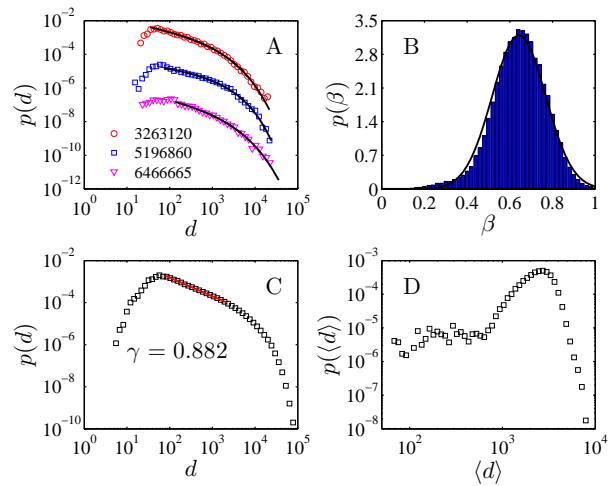}
  \caption{\label{Fig3:WeibullGroup} Classified results of individuals with Weibull distributions of intraday inter-call durations. ({\textit{A}}) Probability distributions of intraday durations for three randomly chosen individuals. The markers of individual 5196860 and 6466665 are shifted vertically by a factor of $10^{-2}$ and $10^{-4}$ for better visibility. The solid lines are the best MLE fit to the Weibull distributions, which give the Weibull exponents $\beta = 0.54$, $\beta = 0.64$, and $\beta = 0.51$ for individual 3263120, 28012863, and 6466665, respectively. ({\textit{B}}) Distribution of the Weibull exponents and the solid curve stands for the fits to normal distribution. ({\textit{C}}) Probability distribution of collective inter-call durations by aggregating the durations of different individuals from the Weibull group as one sample. The solid line is the best fit to the data by means of the least-square method, which gives an estimation of power-law exponent $\gamma = 0.882$. ({\textit{D}}) Probability distribution of the mean inter-call durations for the Weibull group.}
\end{figure}

We investigate the distribution of the aggregate intraday inter-call
durations by treating the individual durations from the power-law group
as one unique sample. To this end, for the aggregate data set in
Fig.~\ref{Fig2:PowerLawGroup}{\textit{C}}, we find a power law with exponent
$\gamma = 1.69$. We find another striking feature in the power-law tail:
the Weibull shape disappears.  Figure~\ref{Fig2:PowerLawGroup}{\textit{D}} plots
the probability distribution of the mean of inter-call durations of the
individuals in the power-law group, where the peak agrees well with the
left peak in Fig.~\ref{Fig1:Agg:Durations}{\textit{D}}.

Figure~\ref{Fig3:WeibullGroup} plots the probability distribution of
intra-call durations for three randomly chosen individuals (3263120,
28012863, and 6466665) whose inter-call durations follow
a Weibull distribution. The solid lines are the best maximum likelihood
estimation (MLE) fits to the Weibull distribution and the corresponding
Weibull exponents are $\beta = 0.54$, $\beta = 0.64$, and $\beta = 0.51$
for individuals 3263120, 28012863, and 6466665, respectively. Having the
Weibull exponents for all individuals from the Weibull group, we
calculate the mean value of the Weibull exponents $\langle \beta \rangle
= 0.64 \pm 0.12$.  Figure~\ref{Fig3:WeibullGroup}{\textit{B}} shows the
distribution of the Weibull exponents $\beta$.  For sake of comparison,
we also present a normal distribution with the parameters obtained by
MLE fits on the sample of Weibull exponents $\beta$. The overlapping
between the empirical data and the normal distribution indicates that
the exponent $\beta$ follows the normal distribution.

Figure~\ref{Fig3:WeibullGroup}{\textit{C}} shows the distribution of inter-call
durations for the Weibull group at the aggregate level. Note that the
functional forms of the distribution in
Fig.~\ref{Fig3:WeibullGroup}{\textit{C}} and the empirical distributions in
Fig.~\ref{Fig1:Agg:Durations}{\textit{A}} are similar, suggesting that the
distributions of the aggregate samples are dominated by individuals
with Weibull duration distributions. Figure~\ref{Fig3:WeibullGroup}{\textit{D}}
plots the probability distribution of the mean inter-call durations for
the individuals in the Weibull group. The peak is in good agreement with
the right peak in Fig.~\ref{Fig1:Agg:Durations}{\textit{D}}.

\begin{figure}[htbp]
 \centering
 \includegraphics[width=7cm]{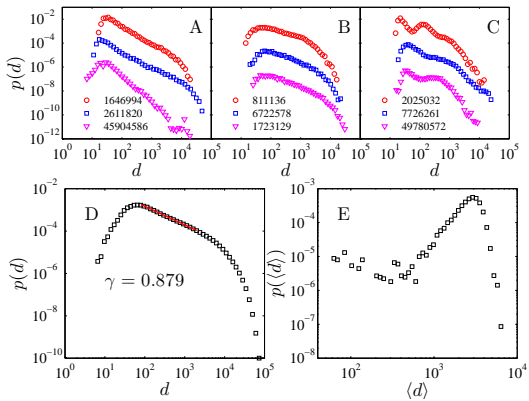}
  \caption{\label{Fig4:OtherGroup} Analysis of the remaining individuals. ({\textit{A}}) Probability distribution of inter-call durations for three individuals, whose durations approach to power-law behaviors without passing the statistical tests. ({\textit{B}}) Probability distribution of inter-call durations for three individuals, whose duration distributions are like Weibull shape but not confirmed by the statistical tests. ({\textit{C}}) Plots of duration distributions for three individuals, whose distribution shapes are uncommon. ({\textit{D}}) Probability distribution of collective inter-call durations by aggregating the durations of different individuals from the Weibull group as one sample. The solid line is the best fit to the data by means of the least-square method, which gives an estimation of power-law exponent $\gamma = 0.879$. ({\textit{E}}) Probability distribution of the mean inter-call durations for the remaining individuals.}
\end{figure}

We next investigate the distribution of inter-call durations for the
remaining 23,197 individuals. We find that for a small fraction of
individuals (close to 2\%), the inter-call durations follow a power law,
as shown in Fig.~\ref{Fig4:OtherGroup}{\textit{A}}.  Because our
statistical tests reject the null hypothesis that individuals follow approximately a
power law, these individuals are excluded from the power-law group. We
find that more than 97\% of the individuals have Weibull tail
distributions, as shown in Fig.~\ref{Fig4:OtherGroup}{\textit{B}}.  However
the fact that the fitting range is lower than 1.5 orders of magnitude
(83\% of the individuals are in the range of $[1, 1.5]$) disallows these
individuals from being classified in the Weibull
group. Figure~\ref{Fig4:OtherGroup}{\textit{C}} shows a very small number
of individuals whose inter-call durations cannot be described by either
power law or Weibull distributions.  Because most of the individuals
have Weibull-tail distributions, the distributions of aggregate
inter-call durations and the mean inter-call durations exhibit patterns
very similar to the results obtained from the Weibull group [see
  Figs.~\ref{Fig4:OtherGroup}{\textit{D}} and
  \ref{Fig4:OtherGroup}{\textit{E}} and Figs.~\ref{Fig3:WeibullGroup}{\textit{C}}
  and \ref{Fig3:WeibullGroup}{\textit{D}}].

{\textbf{Calling patterns for power-law and Weibull groups.}}
Using three measurements, we quantitatively distinguish the calling
patterns of the individuals belonging to two different classified
groups.

\begin{itemize}

\item[{(i)}] The out-degree $k_i$ describes the number of different
  callees for a specified cellphone user.

\item[{(ii)}] The percentage of outgoing calls $r_{\rm{out}}$, is
  defined by dividing the number of outgoing calls by the total number
  of calls---note that the number sending spams (junk message pusher)
  is characterized by $r_{\rm{out}} = 1$.

\item[{(iii)}] The communication diversity $\phi$.  Motivated by the
  social diversity proposed in
  Ref.~\cite{Eagle-Macy-Claxton-2010-Science}, we define the
  communication diversity $\phi_i$, as a function of Shannon entropy to
  quantify how the cellphone users split the number of calls to their
  friends,
  \begin{eqnarray} \label{Eq:Call:Diversity}
  \phi_i = \frac{-\sum_{j=1}^{k_i} p_{ij} \log(p_{ij})}{\log(k_i)}.
  \end{eqnarray}
  Here $k_i$ is the out-degree and $p_{ij}$ is the probability defined
  as $p_{ij}= n_{I}^{ij} / n_{I}^{i} = n_{I}^{ij} / \sum_j^k
  n_{I}^{ij}$, where $n_I^{ij}$ is the number of outgoing calls from
  individual $i$ to individual $j$ and $n_{I}^{i}$ is the total number
  of outgoing calls for individual $i$. A higher $\phi_i$ value
  indicates that the caller's outgoing calls are split more evenly to his
  friends and a smaller $\phi_i$ value implies that most of the
  caller's outgoing calls are to only one of his friends.  Note that we define
  $\phi_i = 0$ when $k_i = 1$.

\end{itemize}

\begin{figure}[htbp]
  \centering
  \includegraphics[width=8cm]{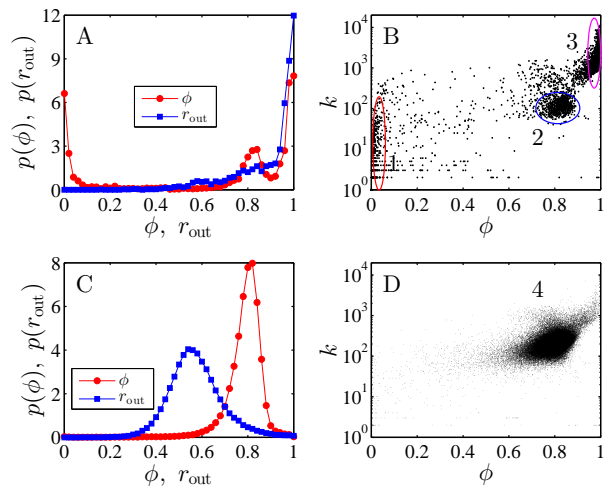}
  \caption{\label{Fig5:Camparison:PL:WBL} Calling patterns for the individuals from power-law and Weibull group. ({\textit{A}}) Distribution of the percentage of outgoing calls $r_{\rm{out}}$ and the call diversity $\phi$ for power law group. ({\textit{B}}) Plots of out-degree $k$ with respect to communication diversity $\phi$ for power law group. Three ellipses correspond to the three clusters of individuals. ({\textit{C}}) Similar as ({\textit{A}}) but for Weibull group. ({\textit{D}}) Similar as ({\textit{B}}) but for Weibull group.}
\end{figure}

In order to distinguish between the calling patterns of the power-law
group of Fig.~\ref{Fig2:PowerLawGroup} and the Weibull group of
Fig.~\ref{Fig3:WeibullGroup}, in
Fig.~\ref{Fig5:Camparison:PL:WBL} we plot the distribution of the
percentage of outgoing calls $r_{\rm{out}}$ and the distribution of the
communication diversity $\phi$.
Figures~\ref{Fig5:Camparison:PL:WBL}{\textit{A}} and
\ref{Fig5:Camparison:PL:WBL}{\textit{C}} compare strikingly different
patterns: (i) in the power-law group, the probability $p(r_{\rm{out}})$
is a monotonically increasing function of $r_{\rm{out}}$ that reaches a
maximum value at $r_{\rm{out}} = 1$ (the characteristic value for spam),
but in the Weibull group, the frequency $p(r_{\rm{out}})$ is a
non-monotonic function of $r_{\rm{out}}$ that has its maximum value
close to the center at $r_{\rm{out}} = 0.56$, and (ii) in the power-law
group, the probability $p(\phi)$ exhibits three pronounced peaks at
$\phi = 0$, $\phi = 0.84$, and $\phi = 1$, but in the Weibull group, the
probability $p(\phi)$ has only one peak at $\phi = 0.82$. We further
estimate the average value of the percentage of outgoing calls $\langle
r_{\rm{out}} \rangle = 0.89 \pm 0.13$ for the power-law group and
$\langle r_{\rm{out}} \rangle = 0.57 \pm 0.11$ for the Weibull
group. Our analysis indicates that the individuals in the power-law
group exhibit more extreme calling behaviors than those in the Weibull
group, e.g., highly-frequent call initiation, a high percentage of
outgoing calls, and either all calls to only one callee or equally
distributing calls among all callees.

Figures \ref{Fig5:Camparison:PL:WBL}{\textit{B}} and
\ref{Fig5:Camparison:PL:WBL}{\textit{D}} plot the out-degree $k$ with
respect to the communication diversity $\phi$ and thus provide
additional evidence that the behavior of individuals in the power-law
group differs greatly from the behavior of individuals in the Weibull
group.  The individuals in the power-law group form three clusters in
the $(\phi, k)$ plane, which are highlighted by the three ellipses in
panel ({\textit{B}}). The three clusters are also consistent with the three peaks
of $f(\phi)$ in Fig.~\ref{Fig5:Camparison:PL:WBL}{\textit{A}}.
Figure~\ref{Fig5:Camparison:PL:WBL}{\textit{D}}, on the other hand, shows
only one large cluster for the Weibull group. Taking the two panels
together, we see that the communication diversity $\phi$ increases
with the out-degree $k$ on average.
In the power-law group we further assign
the individuals with $\phi \le 0.1$ to cluster 1, the individuals with
$0.7 \le \phi \le 0.9$ and $50 \le k \le 200$ to cluster 2, and the
individuals with $\phi \ge 0.9$ and $k \ge 700$ to cluster 3. We find
that there are 762, 710, and 1369 individuals, respectively, with
average degrees of 21.76, 114.98, and 2083.3, respectively, in which the
mean percentage of outgoing calls is 0.99, 0.80, and 0.94 in clusters 1,
2, and 3, respectively. We assign the individuals in the Weibull group
to cluster 4 and find that the average degree and mean percentage of
outgoing calls are 245.13 and 0.57, respectively.  From our analysis, we
first infer that the individuals in power-law cluster 1---the ones
characterized by a high frequency of call initiation, a small number of
callees, or an allocation of almost all outgoing calls to only one
callee---are robot-based users. We next see that the individuals in
cluster 3---the ones characterized by high frequency of call initiation,
a large number of callees, and an even distribution of outgoing calls
among all callees---are associated with telecom frauds and telephone
sales.  We also note that the individuals in cluster 4 are ordinary
cellphone users. We next describe further differences in cellphone
communication activities among the 4 clusters, e.g., the probability
that a caller will call the $c_r$-th most contact and the burst size
probability during burst periods.

\begin{figure}[htbp]
 \centering
 \includegraphics[width=8cm]{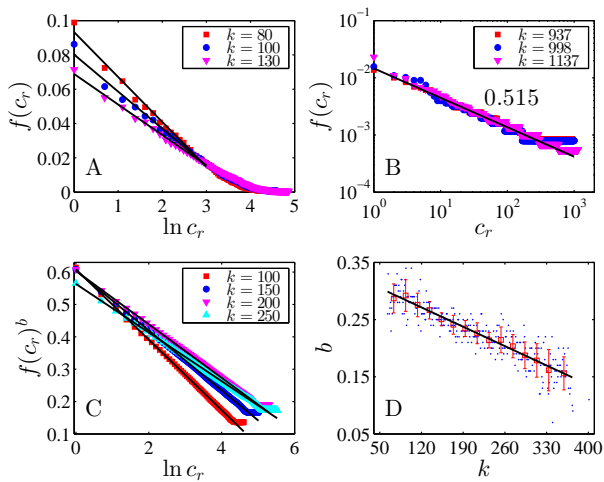}
  \caption{\label{Fig6:MostContacted:PL:WBL} Rank ordering plot showing the average calling frequency $f(c_r)$ of the $c_r$-th most contacted friend for the users with the same degree. ({\textit{A}}) Plots of $f(c_r)$ as a function of $\ln c_r$ for cluster 2. ({\textit{B}}) Loglog plots of $f(c_r)$ with respect to $c_r$ for cluster 3. ({\textit{C}}) Plots of $f(c_r)^b$ versus $\ln c_r$ for cluster 4. ({\textit{D}}) Scatter plots of $b$ with respect to $k$ for cluster 4.}
\end{figure}

Because most of the calls (mean 99.5\% and min 94\%) made by individuals
in cluster 1 are to only one contact, we now calculate the probability
that individuals belonging to the other 3 clusters will only call the
$c_r$-th most contact.  In order to rule out the influence of newly
entering cellphone users, we take into account only those individuals
listed in the data on the starting date of 28 June 2012.
Figure~\ref{Fig6:MostContacted:PL:WBL} shows the average calling
frequency $f(c_r)$ of the $c_r$-th most contact friends for the
individuals with the same degree in cluster 2. There is a linear
relationship between $f(c_r)$ and $\ln c_r$ in panel {\textit{A}}, which
indicates an exponential distribution in the number of outgoing calls to
different contacts \cite{Sornette-2000}.  We see that the slope obtained
between $f(c_r)$ and $\ln c_r$ increases as the out-degree $k$
increases, but the lack of individuals prevents us from finding the
functional form between the slopes and the out-degree values $k$.  We
also observe power-law behavior between $f(c_r)$ and $c_r$ in cluster 3
of panel ({\textit{B}}).  The least-square linear fits provide an estimate for
power-law exponent $0.52$ and also show that the behavior of the
power-law exponent is not affected by the out-degree $k$.
Figure~\ref{Fig6:MostContacted:PL:WBL}{\textit{C}} plots $f(c_r)^b$ versus
$\ln c_r$ for cluster 4, where $b$ is associated with the maximum
correlation coefficient of least-square linear fits to $f(c_r)^b$ versus
$\ln c_r$ by varying $b$ from 0.01 to 0.99 with a step of 0.01.  The
linear relationship between $f(c_r)^b$ and $\ln c_r$ suggests that the
number of outgoing calls to contacts follows a stretched exponential
distribution \cite{Sornette-2000,Laherrere-Sornette-1998-EPJB}.  In
panel ({\textit{D}}) we show the exponent $b$ plotted with respect to the
out-degree $k$, where we observe a striking linear relationship: $b =
-4.836\times10^{-4}k+0.329$. Here we report that the probability to call
the $c_r$-th most contact is in sharp contrast to the results reported in
Ref.~\cite{Bagrow-Lin-2012-PLoS1}, where a Zipf law with a power-law
exponent 1.5 is observed when, in contrast to our ``microscopic'' study,
individuals are not grouped according to their distributions of
inter-call durations.

It was recently proposed that the distribution of burst sizes indicates
the presence of memory behaviors in the timing of consecutive events
\cite{Karsai-Kaski-Barabasi-Kertesz-2012-SR}, where the deviation from
exponential distributions is a hallmark of correlated properties. For a
given series of events, a burst period is a cluster of consecutive
events following their previous events within a short time interval $\Delta t$,
which is an arbitrarily
assigned value in empirical analysis. The burst size $e_b$ is defined as
the number of events in a burst period.

\begin{figure}[htbp]
  \centering
  \includegraphics[width=8cm]{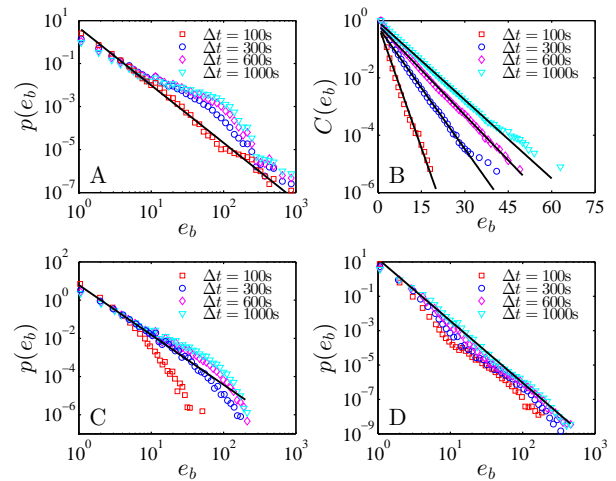}
  \caption{\label{FFig7:NumEvents:PL:WBL} Distribution of burst sizes $e_b$ in burst periods. ({\textit{A}}) Cluster 1 (PDF). ({\textit{B}}) Cluster 2 (CDF). ({\textit{C}}) Cluster 3 (PDF). ({\textit{D}}) Cluster 4 (PDF). }
\end{figure}

Based on our dataset, Fig.~\ref{FFig7:NumEvents:PL:WBL}
shows the probability distribution of burst sizes $e_b$ in burst periods
at the aggregate level for the 4 clusters by setting $\Delta t$ = 100,
300, 600, and 1000 seconds.  In panel {\textit{A}} we find the probability
distributions of $e_b$ for cluster 1.  Although we find a very good
power-law relationship between $p(e_b)$ and $e_b$ with an exponent 2.677
for $\Delta t = 100$ seconds, the distributions deviate from power-law
distributions and tend to exponential distributions for $\Delta t$ =
300, 600, and 1000 seconds. Panel ({\textit{B}}) shows that the probability
distribution of $e_b$ exhibits excellent exponential distributions for
varying values of $\Delta t$ for cluster 2. Panel ({\textit{C}}) plots the
probability distributions of $e_b$ for cluster 3. We see the power-law
behavior of $p(e_b)$ with an exponent 2.61 only when $\Delta t = 300$
seconds. When $\Delta t$ = 600 and 1000 seconds, $p(e_b)$ switches from
power-law behavior to a bimodal pattern (with exponential tails). Panel
({\textit{D}}) shows that the probability distributions of $e_b$ corresponding to
different values of $\Delta t$ for individuals in cluster 4 all display
very good power-law behavior, and that the power-law decay exponent is
3.6.  Comparing our distribution with the distribution reported in
Fig.~2{\textit{A}} in Ref.~\cite{Karsai-Kaski-Barabasi-Kertesz-2012-SR}, we find
that the distribution shapes are very similar for cluster 4, the only
difference being that the extremely large brust sizes $e_b \ge 500$
disappear in the plots for the individuals with very long burst sizes
assigned into cluster 1.

\bigskip
{\textbf{Discussion}}

Contrary to common belief, we find that only 3.46\% of callers have
inter-call durations that follow a power-law distribution. The majority
of callers (73.34\%) have inter-call durations that follow a Weibull
distribution. Further examination reveals that callers with a power-law
distribution exhibit anomalous and extreme calling patterns often linked
to robot-based calls, telecom frauds, or telephone sales---information
valuable to both academics and practitioners, especially mobile telecom
providers. We note that Weibull distributions are ubiquitous in such 
routine human activities as intervals for online gamers \cite{Jiang-Zhou-Tan-2009-EPL} 
and intertrade intervals in stock trading \cite{Ivanov-Yuen-Podobnik-Lee-2004-PRE,Jiang-Chen-Zhou-2008-PA}.

Although most of the individuals exhibit Weibull distributions of the inter-call durations, the distribution at the population level is a power law with an exponential cutoff, consistent with other works using mobile phone communication data from other sources \cite{Gonzalez-Hidalgo-Barabasi-2008-Nature,Karsai-Kaski-Barabasi-Kertesz-2012-SR}. We argue that a superposition of individuals' heterogeneous calling behaviors leads to the exponentially truncated power-law distribution at the population level, showing the importance of different characteristic scales.

Although individual callers exhibit heterogeneities across the entire
population and their personal activities are also heterogeneous,
individual callers can be grouped into clusters according to their
similarities. The findings reported in this paper enable us to
construct dynamic models at an individual level that agree with
empirical collective properties. Every reasonable dynamic model for
cellphone usage should include the major findings of this paper, i.e.,
that individuals are not identical and do not exhibit identical
behavior. Our strategy is to propose models based, not on individuals,
but on clusters of individuals. Thus to accurately model the trigger
process in human activity we need a precise classification of
individuals according to the similarities in their activities, and also
a detailed investigation of the complete activity log for each
individual.

\bigskip
{\textbf{Materials and Methods}}

{\textbf{Data description.}}
Our data, which are provided by a cellphone provider in China, contain
all the calling records covering two periods. One is from 28 June 2010
to 24 July 2010 and the other is from 1 October 2010 to 31 December
2010. For unknown reasons, the calling logs for a few hours on certain
days (Ocotober 12, November 5, 6, 13, 21, and 27 and December 6, 8, 21,
and 22) are missing, and they are excluded from our analysis, which
results in a total of 109 days.

\begin{figure}[htbp]
 \centering
  \includegraphics[width=8cm]{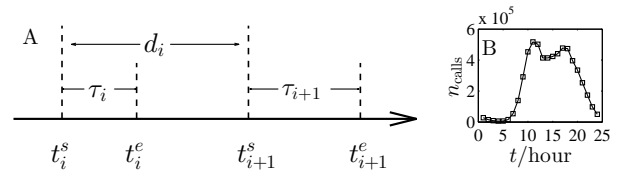}
  \caption{\label{Fig8:Definition} Definition of intraday inter-call durations. ({\textit{A}}) Schematic chart of call logs for an individual. ({\textit{B}}) Intraday pattern of the number of calls.}
\end{figure}

For each entry of record, we have the information of caller number,
callee number, call starting time, call length, and call status. The
caller and callee number is encrypted in order to protect personal
privacy. The call status indicates whether the call is terminated
normally. Note that we only take into account normal calls that the
call begins and ends normally. The calls that are not completed or are
interrupted, are also discarded. To better explain our data, Fig.~\ref{Fig8:Definition} {\textit{A}} 
shows the call records for a given individual subscriber, where a call starts at $t^s$ and ends at $t^e$. We usually have
\begin{equation}
 \cdots < t_i^s < t_i^e < t_{i+1}^s < t_{i+1}^e < \cdots.
 \label{Eq:Data:CE}
\end{equation}
Further examination is made to check whether $t_i^e$ is less than
$t_{i+1}^s$ for each individual. The records that do not obey the
equation $t_i^e < t_{i+1}^s$ can be attributed to the recording errors
introduced by the system and the $i+1$-th call record is discarded.

{\textbf{Definition of intraday inter-call durations.}}
As shown in Fig.~\ref{Fig8:Definition} {\textit{A}}, the inter-call duration is defined as the time that elapses between two
consecutive calls and it can be calculated via $d_i = t^s_{i} -
t^s_{i-1}$. In order to avoid the influence on the results of
discontinuous recording days, which produce very large inter-call
durations, we restrict the durations to a period of one day (the typical
human circadian rhythm).  Although it might seem obvious to separate the
days at midnight (00:00 AM), late night calls (made by lonely people,
lovers, and friends) are common, so we divide the days at 4:00 A.M.,
which is the time point associating with the lowest call volume in a
24-hour period [see Fig.~\ref{Fig8:Definition}{\textit{B}}]. This allows us to
take into account the people who go out and stay awake later as
well. Our restriction is equivalent to excluding inter-call durations
that span the dividing point (4:00 AM).

{\textbf{Fitting distributions and statistical tests.}}
A simple approach based on maximum likelihood estimation (MLE) fits and
Kolmogorov-Smirnov (KS) tests is used to check whether the candidate
distributions (power-law or Weibull) can be used to fit the individual
intraday inter-call durations. Because people are more interested in
the distribution form of large durations, we assume that the durations
larger than a truncated value $d_{\min}$ are described by the candidate
distributions, such that
\begin{eqnarray}
 p(d) & \sim & d^{-\gamma}, ~~d \ge d_{\min} \\
 p(d) & = & \alpha \beta d^{\beta-1} \exp(-\alpha d^\beta), ~~ d \ge d_{\min}.
\end{eqnarray}
We also determine the lowest boundary $d_{\min}$ as an additional
parameter.  Once $d_{\min}$ is obtained, the distribution parameters can
be estimated by means of MLE fits to the left-truncated candidate
distribution. Hence, the accuracy of estimated $d_{\min}$ plays an
important role in estimating accurate distribution parameters. Inspired
by the method proposed in Ref.~\cite{Clauset-Shalizi-Newman-2009-SIAMR},
the best $d_{\min}$ is associated with the truncated sample with the
smallest KS value. The truncated sample is obtained by discarding the
durations below $d_{\min}$ in the original duration sample. After the
lowest $d_{\min}$ and the corresponding distribution parameters are
obtained, we use the KS test and CvM test to check the fitting. The null
hypothesis $H_0$ for our KS test and CvM test is that the data ($d >
d_{\min}$) are drawn from the candidate distribution (power-law
distribution or Weibull distribution).

{\textbf{Determining the distribution form.}}
The sample of individual intraday inter-call durations, which we assume
conforms to a power-law distribution, must (i) pass either of the two
tests at the significant level 0.01 and (ii) exhibit a fitting range of
not less than 1.5 orders of magnitude. For Weibull distributions, in
addition to the two above conditions, the Weibull exponent $\beta$ of
the intraday duration sample must be in the range $(0, 1)$. Because a
power-law distribution is a two-parameter model and a Weibull
distribution is a three-parameter model, we first filter out the
individuals with durations that follow a power-law distribution and than
inject the remaining individuals into the Weibull filtering procedure.

\bigskip
{\textbf{Acknowledgments:}}

Z.-Q.J., W.-J.X., M.-X.L. and W.-X.Z received support from the National Natural Science
Foundation of China Grant 11205057, the Humanities and Social
Sciences Fund (Ministry of Education of China Grant 09YJCZH040),
and the Fok Ying Tong Education Foundation Grant 132013. BP and
HES received support from the Defense Threat Reduction Agency (DTRA), the Office of Naval Research (ONR), and the National Science Foundation (NSF) Grant CMMI 1125290.



\begin{thebibliography}{10}

\bibitem{Eckmann-Moses-Sergi-2004-PNAS}
Eckmann JP, Moses E, Sergi D (2004) {Entropy of dialogues creates coherent
  structures in e-mail traffic}. \emph{Proc Natl Acad Sci USA}
  101:14333--14337.

\bibitem{Palla-Barabasi-Vicsek-2007-Nature}
Palla G, Barab{\'a}si AL, Vicsek T (2007) {Quantifying social group evolution}.
  \emph{Nature} 446:664--667.

\bibitem{Onnela-Saramaki-Hyvonen-Szabo-Menezes-Kaski-Barabasi-Kertesz-2007-NJP}
Onnela JP, \emph{et~al.} (2009) {Analysis of a large-scale weighted network of
  one-to-one human communication}. \emph{New J Phys} 9:179.

\bibitem{Onnela-Saramaki-Hyvonen-Szabo-Lazer-Kaski-Kertesz-Barabasi-2007-PNAS}
Onnela JP, \emph{et~al.} (2007) {Structure and tie strengths in mobile
  communication networks}. \emph{Proc Natl Acad Sci USA} 104:7332--7336.

\bibitem{Jo-Pan-Kaski-2011-PLoS1}
Jo HH, Pan RK, Kaski K (2011) {Emergence of bursts and communities in evolving
  weighted networks}. \emph{PLoS One} 6:e22687.

\bibitem{Kumpula-Onnela-Saramaki-Kaski-Kertesz-2007-PRL}
Kumpula JM, Onnela JP, Saram{\"a}ki J, Kaski K, Kert{\'e}sz J (2007) {Emergence
  of communities in weighted networks}. \emph{Phys Rev Lett} 99:228701.

\bibitem{Barabasi-2005-Nature}
Barab{\'a}si AL (2005) {The origin of bursts and heavy tails in human
  dynamics}. \emph{Nature} 435:207--211.

\bibitem{Holme-Saramaki-2012-PR}
Holme P, Saram{\"a}ki J (2012) {Temporal networks}. \emph{Phys Rep}
  519:97--125.

\bibitem{Pan-Saramaki-2011-PRE}
Pan RK, Saram{\"a}ki J (2011) {Path lengths, correlations, and centrality in
  temporal networks}. \emph{Phys Rev E} 84:016105.

\bibitem{Malmgren-Stouffer-Motter-Amaral-2008-PNAS}
Malmgren RD, Stouffer DB, Motter AE, Amaral LAN (2008) {A Poissonian
  explanation for heavy tails in e-mail communication}. \emph{Proc Natl Acad
  Sci USA} 105:18153--18158.

\bibitem{Hong-Han-Zhou-Wang-2009-CPL}
Hong W, Han XP, Zhou T, Wang BH (2009) {Heavy-tailed statistics in
  short-message communication}. \emph{Chin Phys Lett} 26:028902.

\bibitem{Wu-Zhou-Xiao-Kurths-Schellnhuber-2010-PNAS}
Wu Y, Zhou CS, Xiao JH, Kurths J, Schellnhuber HJ (2010) {Evidence for a
  bimodal distribution in human communication}. \emph{Proc Natl Acad Sci USA}
  107:18803--18808.

\bibitem{Zhao-Xia-Shang-Zhou-2011-CPL}
Zhao ZD, Xia H, Shang MS, Zhou T (2011) {Empirical analysis on the human
  dynamics of a large-scale short message communication system}. \emph{Chin
  Phys Lett} 28:068901.

\bibitem{Candia-Gonzalez-Wang-Schoenharl-Madey-Barabasi-2008-JPAMT}
Candia J, \emph{et~al.} (2008) {Uncovering individual and collective human
  dynamics from mobile phone records}. \emph{J Phys A: Math Theor} 41:224015.

\bibitem{Karsai-Kaski-Barabasi-Kertesz-2012-SR}
Karsai M, Kaski K, Barab{\'a}si AL, Kert{\'e}sz J (2012) {Universal features of
  correlated bursty behaviour}. \emph{Sci Rep} 2:397.

\bibitem{Oliveira-Barabasi-2005-Nature}
Oliveira JG, Barab{\'a}si AL (2005) {Darwin and Einstein correspondence
  patterns}. \emph{Nature} 437:1251.

\bibitem{Li-Zhang-Zhou-2008-PA}
Li NN, Zhang N, Zhou T (2008) {Empirical analysis on temporal statistics of
  human correspondence patterns}. \emph{Physica A} 387:6391--6394.

\bibitem{Malmgren-Stouffer-Campanharo-Amaral-2009-Science}
Malmgren RD, Stouffer DB, Campanharo ASLO, Amaral LAN (2009) {On Universality
  in Human Correspondence Activity}. \emph{Science} 325:1696--1700.

\bibitem{Karsai-Kaski-Kertesz-2012-PLoS1}
Karsai M, Kaski K, Kert{\'e}tz J (2012) {Correlated dynamics in egocentric
  communication networks}. \emph{PLoS One} 7:e40612.

\bibitem{Gonzalez-Hidalgo-Barabasi-2008-Nature}
Gonz{\'a}lez MC, Hidalgo CA, Barab{\'a}si AL (2008) {Understanding individual
  human mobility patterns}. \emph{Nature} 453:779--782.

\bibitem{Clauset-Shalizi-Newman-2009-SIAMR}
Clauset A, Shalizi CR, Newman MEJ (2009) {Power-law distributions in empirical
  data}. \emph{SIAM Rev} 51:661--703.

\bibitem{Eagle-Macy-Claxton-2010-Science}
Eagle N, Macy M, Claxton R (2010) {Network diversity and economic development}.
  \emph{Science} 328:1029--1031.

\bibitem{Sornette-2000}
Sornette D (2000) \emph{{Critical Phenomena in Natural Sciences - Chaos,
  Fractals, Self-organization and Disorder: Concepts and Tools}} (Springer,
  Berlin), 1 edn.

\bibitem{Laherrere-Sornette-1998-EPJB}
Laherr{\`e}re J, Sornette D (1998) {Stretched exponential distributions in
  nature and economy: ``Fat tails'' with characteristic scales}. \emph{Eur Phys
  J B} 2:525--539.

\bibitem{Bagrow-Lin-2012-PLoS1}
Bagrow JP, Lin YR (2012) {Mesoscopic structure and social aspects of human
  mobility}. \emph{PLoS One} 7:e37676.

\bibitem{Jiang-Zhou-Tan-2009-EPL}
Jiang ZQ, Zhou WX, Tan QZ (2009) {Online-offline activities and game-playing
  behaviors of avatars in a massive multiplayer online role-playing game}.
  \emph{EPL (Europhys Lett)} 88:48007.

\bibitem{Ivanov-Yuen-Podobnik-Lee-2004-PRE}
Ivanov PC, Yuen A, Podobnik B, Lee YK (2004) {Common scaling patterns in
  intertrade times of U. S. stocks}. \emph{Phys Rev E} 69:056107.

\bibitem{Jiang-Chen-Zhou-2008-PA}
Jiang ZQ, Chen W, Zhou WX (2008) {Scaling in the distribution of intertrade
  durations of Chinese stocks}. \emph{Physica A} 387:5818--5825.

\end{thebibliography}

\end{document}